\documentclass[apj]{emulateapj}
\usepackage{xspace}
\journalinfo{Accepted for publication in the Astronomical Journal}
\slugcomment{To appear in the Astronomical Journal in October 2004}

\newcommand\msun{\ensuremath{M_\sun}\xspace}
\newcommand\mcrit{\ensuremath{M_{crit}}\xspace}

\shorttitle{White dwarf candidates in Praesepe}
\shortauthors{Williams, Bolte, \& Liebert}

\begin{document}

\title{Spectroscopic Identification of Faint White Dwarf Candidates in
  the Praesepe Open Star Cluster} 
\author{Kurtis A. Williams}
\affil{Steward Observatory \\ 933 N. Cherry Ave, Tucson, AZ 85721}
\email{kurtis@as.arizona.edu}
\author{Michael Bolte}
\affil{UCO/Lick Observatory \\ University of California, Santa Cruz, CA 95064}
\email{bolte@ucolick.org}
\and
\author{James W. Liebert}
\affil{Steward Observatory \\ 933 N. Cherry Ave., Tucson, AZ 85721}
\email{jliebert@as.arizona.edu}

\begin{abstract}
We present spectroscopic observations of the remaining four candidate
white dwarfs in Praesepe.  All four candidates are quasars with
redshifts between 0.8 and 2.8. One quasar, LB 6072, is observed to
have a strong metal-line absorption system blueward of the quasar
redshift.  The lack of additional white dwarfs in Praesepe leaves the
total known white-dwarf population of the cluster at five, well below the
number expected from commonly-assumed initial mass functions, though
several undiscovered cluster WDs may lie in the outer regions of the
cluster. All known
Praesepe member white dwarfs are concentrated within $0\,\fdg 6$  of the
cluster center, and the radial profile of cluster white dwarfs is
quite similar to 
the profile of massive cluster stars. This profile is mildly
inconsistent with that of $\sim 1\msun$ cluster stars and suggests
that the white dwarfs
did not receive a velocity kick during the progenitor star's mass loss
phases.  If complete, the observed Praesepe white dwarf
population is consistent with a steeper high-end initial-mass function
than commonly assumed, though the calculated slopes are inconsistent
with the present-day mass function of Praesepe.  Searches for white
dwarfs outside the core of Praesepe and further study of the
white dwarf populations of additional open clusters is necessary to
constrain further the underlying cause of the white dwarf deficit.
\end{abstract}

\section{Introduction}
White dwarfs (WDs) are the final stage of stellar evolution for the
vast majority of stars.  WDs are known to exist in open clusters with
main-sequence turnoff masses of $\sim 5\msun$
\citep[e.g. \objectname{NGC 2516},][]{Koester1996} and may have
progenitors with zero-age main sequence masses as large as $8\msun$.
However, the WD population of the \objectname{Hyades} shows a deficit
of WDs compared with the number that would be expected given the
present-day mass function and reasonable assumptions for the shape of
the initial-mass function (IMF) \citep{Weidemann1992}, and no Hyades
WDs have inferred progenitor masses $\gtrsim 3.6\msun$ \citep[Claver
et al. 2001, hereafter ][]{Claver2001}.

There are at least three explanations for this white dwarf deficit.
First, the ``missing'' WDs may be hidden in unresolved binary star
systems. A calculation of the number of hidden WDs finds that this can
explain many, but not all, of the missing WDs  \citep[Williams 2004,
hereafter ][]{Williams2004}. Second, dynamical evolution of the open
cluster may remove WDs from the cluster \citep{Weidemann1992}.  This
effect is enhanced if the WDs receive a velocity ``kick'' during the
mass-loss phase of the progenitor star \citep{Fellhauer2003}.  Third,
the higher-mass end of the initial-mass function (IMF) in the Hyades
may have been steeper than commonly-observed IMFs, lacking stars with
$M\gtrsim 4\msun$.

In part to address this issue, \citet{Claver2001} studied the WD
population of the Praesepe open cluster.  Praesepe has an age ($\sim
625$ Myr) and metallicity ($Z=0.024$)  indistinguishable from that of
the Hyades \citep{Claver2001}.  Praesepe has a total mass of $\sim
600\msun$ and a half-mass radius $\approx 3.9$ pc \citep{Adams2002},
more massive and more compact than the Hyades, which has a total mass
$\sim 400\msun$ and a half-mass radius of $\approx 5.7$ pc
\citep{Perryman1998},  so losses of WDs due to dynamical evolution
should be less in Praesepe than in the Hyades.

In Praesepe, \citet{Claver2001} found one WD with a massive progenitor,
  \objectname{LB 1847} = \objectname{EG 60}, $M_i=4.17\msun$, and four
  WDs with less-massive progenitors.  Four other Praesepe WD
  candidates were not observed spectroscopically in that study due to
  their faintness; two of these (\objectname{LB 1839} and
  \objectname{LB 6072}) have photometric properties consistent with
  cluster membership, a large WD mass and a progenitor mass $\gtrsim
  5\msun$. If these objects are indeed cluster member WDs, they would
  be useful objects for exploring the value of \mcrit and would mostly
  resolve the Praesepe WD deficit at these higher masses
  \citep{Williams2004}.

As part of our ongoing study of open cluster WDs, we have obtained
spectra of these four objects.  \S 2 details the observations and
spectroscopic identifications of each object, and \S 3 discusses the
implications of this study.

\section{Observations and data reduction}

Spectra of Praesepe WD candidates were obtained between 2003
and 2004 with the Keck I 10m telescope and the Magellan Baade 6.5m
telescope.  An observing log is presented in Table
\ref{obslog.tab}. We note that the coordinates for \objectname{LB
6037}, \objectname{LB 6072}, \objectname{LB 1839}, and \objectname{LB
1876}  given in Table 3 of \citet{Claver2001} are labeled as J2000
coordinates, though the B1950 coordinates are published.  For the sake of
clarity, J2000 coordinates for  all the objects in \citet{Claver2001}
are given in Table \ref{ids.tab}.

\begin{deluxetable}{lllc}
\tablecolumns{4}
\tablecaption{Observing log.\label{obslog.tab}}
\tablewidth{0pt}
\tablehead{\colhead{UT Date} & \colhead{Telescope} & \colhead{Object}
  & \colhead{Exp time (s)}}
\startdata
08 Dec 2002 & Keck & WD 0834+209 & 600 \\
            &      & LB 1839     & 600 \\ 
12 Feb 2004 & Keck & LB 6072     & 600 \\
22 Mar 2004 & Magellan & LB 6037 & 600 \\
\enddata
\end{deluxetable}

\begin{deluxetable*}{llcccccc}
\tablecolumns{8}
\tablecaption{Spectroscopic Identifications \label{ids.tab}}
\tablewidth{0pt}
\tablehead{\colhead{Object} & \colhead{Other Names} & \colhead{RA
    (J2000)} & \colhead{Dec (J2000)} & \colhead{$V$\tablenotemark{a}}
  & \colhead{\bv\tablenotemark{a}} & \colhead{ID} &
  \colhead{$z\tablenotemark{b}$}} 
\startdata
\objectname{QSO J0837+2040} & \objectname{WD 0834+209}\tablenotemark{a} & 
    8 37 11.5 & 20 40 57.1 & 18.83 & 0.34 & QSO & $1.0916\pm0.0013$ \\ 
\objectname{QSO J0838+1933} & \objectname{LB 1839} &
    8 38 37.2 & 19 33 13.2 & 18.83 &  0.23 & QSO & $0.8622\pm0.0009$ \\
\objectname{QSO J0843+1955} & \objectname{LB 6037} &
    8 43 13.3 & 19 55 59\phantom{.0} & 18.98 &  0.37 & QSO & $2.78\pm0.01$\\
\objectname{QSO J0843+1937} & \objectname{LB 6072} &
    8 43 45.7 & 19 37 24.2 & 18.73 & 0.24 & QSO & $1.9749\pm0.0004$\tablenotemark{c} \\
\objectname{WD 0836+197} & \objectname{LB 5893}    &
    8 39 36.5 & 19 30 43.2 & 17.63 & -0.06 & WD & \nodata \\
\objectname{WD 0836+201} & \objectname{LB 393}, \objectname{EG 61}  &
    8 39 45.6 & 20 00 15.6 & 17.96 &  0.12 & WD & \nodata \\
\objectname{WD 0836+199} & \objectname{LB 1847}, \objectname{EG 60} &
    8 39 47.2 & 19 46 11.8 & 18.33 &  0.08 & WD & \nodata \\ 
\objectname{WD 0837+199} & \objectname{LB 390}, \objectname{EG 59}  &
    8 40 28.1 & 19 43 34.4 & 17.60 & 0.04 & WD & \nodata \\
\objectname{WD 0838+202} & \nodata        &
    8 41 39.8 & 19 00 07.5 & 17.96 &  0.20 & WD & \nodata \\
\objectname{WD 0840+200}\tablenotemark{d} & \objectname{LB 1876} &
    8 42 52\phantom{.0} & 19 51 12\phantom{.0} & 17.69 & 0.15 & WD & \nodata \\
\enddata
\tablenotetext{a}{As given in \citet{Claver2001}, should not be used}
\tablenotetext{b}{$1\sigma$ errors calculated in \emph{fxcor}}
\tablenotetext{c}{Strong metal absorption line system at $z=1.82$}
\tablenotetext{d}{Incorrectly listed in \citet{Claver2001} Table 3 as ``WD 0837+202''}
\end{deluxetable*}

The Keck observations used the upgraded blue camera of the LRIS
spectrograph \citep{Oke1995} with the 400 l mm$^{-1}$, 3400\AA~ blaze
grism and a 1\arcsec-wide longslit at parallactic angle.  The Magellan
observation used the IMACS spectrograph with the f/2 camera, the 200 l
mm$^{-1}$, 6600\AA~ blaze grism, and the 1\arcsec-wide slit in
slit-viewing mode at parallactic angle.  The resulting spectral
resolution was 6\AA ~for the Keck data and 7\AA ~for the Magellan
data.   The spectra were reduced using standard NOAO \emph{IRAF}
routines.  Relative flux calibration was obtained by applying the
response function obtained from observations of spectrophotometric
standards obtained on the corresponding observing run.

\begin{figure}
\plotone{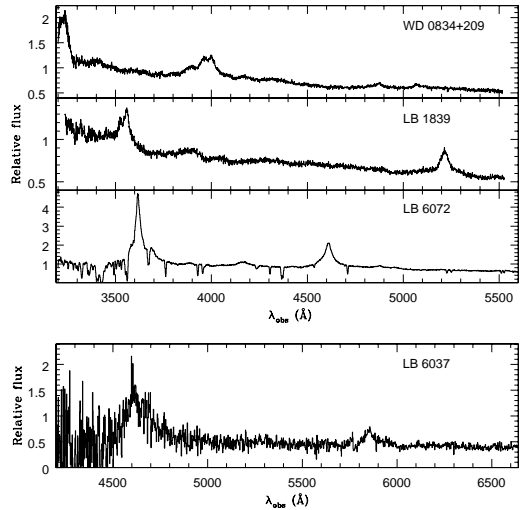}
\caption{Spectra of the four candidate Praesepe white
dwarfs. Spectra are unbinned and in (relative) units of
$f_\lambda$. \label{spectra.fig}} 
\end{figure}

\begin{figure}
\plotone{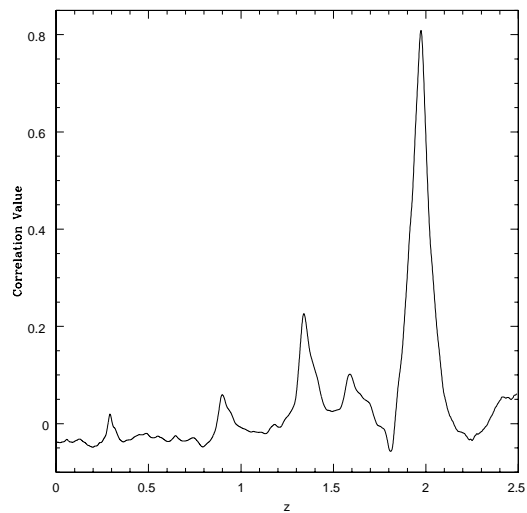}
\caption{Correlation values for LB 6072.  The correlation value as a
  function of redshift for cross-correlation of LB 6072 with the SDSS
  composite quasar spectrum from \citet{Vanden Berk2001}. The sharp,
  high peak is the QSO redshift, $z=1.975$.
\label{correlate.fig}}
\end{figure}

Spectra for each of the four remaining white-dwarf candidates are
presented in Fig.~\ref{spectra.fig}.  As can be seen, none of the four
objects are WDs; all are QSOs.  Redshifts for the QSOs were determined
by cross-correlating the object spectrum with the SDSS composite QSO
spectrum of \citet{Vanden Berk2001} using the \emph{fxcor} package in
\emph{IRAF}. For illustrative purposes, an example of the
cross-correlation values for LB 6072 are shown
Fig.~\ref{correlate.fig}.  The spectral identifications and redshifts
of each object are summarized in Table \ref{ids.tab}. Each of these quasars is
also assigned a QSO catalog designation (listed in the first column
of the table).

\begin{figure}
\plotone{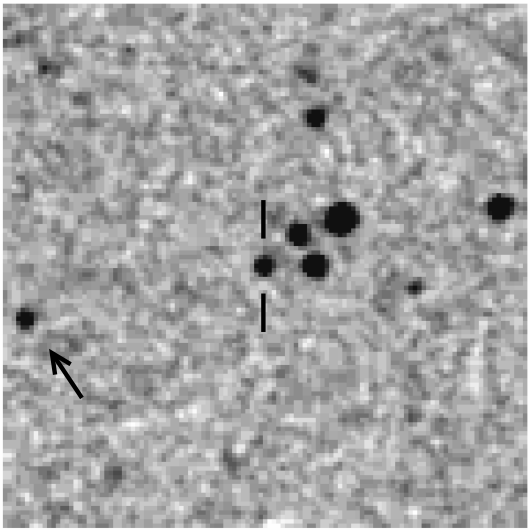}
\caption{Finder chart for LB 6037, indicated by
vertical tick marks.  The arrow indicates a K star at the
coordinates given for LB 6037 in \citet{Wagner1986}.  The image is a
$2\arcmin\times 2\arcmin$ field taken from the blue plate of the
second-generation Digitized Sky Survey.\label{lb6037.fig}}
\end{figure}

Special difficulties were posed by \objectname{LB 6037}.  The
coordinates for this object in SIMBAD and \citet{Claver2001} are from
\cite{Wagner1986}; however, the object at these coordinates is not a
blue object but a K star.  In order to find any nearby blue objects, 
we blinked the blue and red
second-generation Digitized Sky Survey plates from the ESO Online DSS
server.  A point source $\sim 1\arcmin$ west of the given coordinates
stands out as being far bluer than anything else in the field. A
spectrum of the blue source obtained at Magellan reveals it to be a
QSO.  We therefore identify this blue object as LB 6037 and give these
coordinates in Table \ref{ids.tab}.  A finding chart is presented
in Fig.~\ref{lb6037.fig}. Due to the lower signal-to-noise of this
spectrum, the cross-correlation routine failed to converge
satisfactorily.  Manual line identifications give a redshift
of $z=2.78$.

\objectname{LB 1839} was identified by \citet{Claver2001} as a proper
motion member of Praesepe, though it is actually a QSO.  As quasars
have zero proper motion, this is an obvious error in the proper motion
measurement.  Measurements of ``significant'' proper motions for QSOs
have been noted in a cross-correlation of the USNO-B and SDSS catalogs
\citep{Gould2004} and merely illustrate the difficulty in measuring
small proper motions of faint objects.

\section{Discussion}

\subsection{Praesepe WD candidates and sample completeness}

Any conclusions about the Praesepe WD population will depend on the 
completeness of the WD candidate lists. CLBK01 do not discuss the 
completeness of their sample. They also  included a few previously 
known WD candidates outside their survey area (outlined in Figure 4). 
However, the CLBK01 photometry is complete to $V\sim 19.5$, more than 1 
mag fainter than the faintest Praesepe WD, so it is not unreasonable to 
assume that the sample is complete (aside from the complication of WDs 
in binary systems) in the central 2.1 square degrees ($\sim 50^\prime$ 
radius) of Praesepe.

\begin{figure}
\plotone{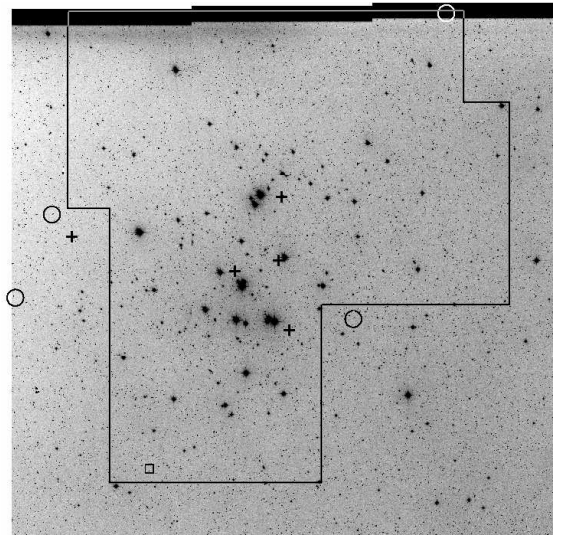}
\caption{Mosaicked Digitized Sky Survey image of
Praesepe. Crosses indicate locations of Praesepe WDs from
\citet{Claver2001}; the square indicates the \citet{Claver2001}
non-Praesepe WD; open circles mark the location of the QSOs from this
study.  The outlined region represents the approximate coverage of
\citet{Claver2001}. The image is $\sim 2\degr$ on a
side. \label{praesepe.fig}}
\end{figure}

\subsection{The Deficit of Praesepe WDs}
 
Members of Praesepe are known to exist out to a radius of $\sim 8\degr$
\citep{Adams2002}, so it is quite possible that more cluster WDs exist
in the outer regions of the cluster, just as the WD population in
the open cluster M67 is observed to extend to the largest cluster
radii \citep{Richer98}.  The potential number of Praesepe WDs outside
the surveyed area can be
estimated by several means.  Using the parameters of the best-fitting
King model determined by \citet{Adams2002}, we find that $\sim 85\%$
of the stellar content of Praesepe is outside the central
$50\arcmin$.  Using the profile of $\sim 1\msun$ stars in
\citet{Raboud1998b}, we estimate that $\sim 80\%$ of the stellar
content is outside the surveyed regions.  Therefore, \emph{if} the WD
population of Praesepe has the same radial distribution as that of the
majority of the stellar population (discussed in \S\ref{sec.kicks}), 
Praesepe may have a total population of $\sim 25$ to 30 WDs. 

The number of expected, observable WDs in Praesepe was calculated in
\citet{Williams2004} to be seven to 21 WDs.  This calculation used the
luminosity function of \citet{Jones91}, which covered the central
$\sim 16$ sq.~degrees of Praesepe.  The large range in the number of
expected WDs arises primarily from the assumed IMF.  If this
calculation is scaled by the radial profile of \citet{Raboud1998b}
under the assumption that the WD radial profile follows that of the
$\sim 1\msun$ stars, we estimate that $\sim 1$ to 5 WDs should be
found in the central region studied by \citet{Claver2001}, in
agreement with the five known Praesepe WDs. If, however, these five
known Praesepe WDs represent the \emph{entire} WD population of the
cluster,  then Praesepe suffers from the same
deficit of WDs as the Hyades \citep{Weidemann1992}.  

\subsubsection{WD Velocity Kicks\label{sec.kicks}}
One explanation for the deficit of WDs in open clusters is that during
the mass-loss phase of the progenitor star's stellar evolution, the
future WD is accelerated due to asymmetric mass loss, resulting in a
small net velocity ``kick''  \citep{Weidemann1992}. A net velocity of
just a few km/s is sufficient to remove the majority of WDs from an
open cluster on short time scales \citep{Fellhauer2003}.  In N-body
simulations not considering velocity kicks, the WD populations of open
clusters with ages similar to Praesepe have radial profiles very
similar to those of the luminous stars and subgiants \citep{Portegies
  Zwart2001,Baumgardt2003}. 

Figure~\ref{praesepe.fig} shows the location of the five Praesepe WD
candidates with respect to the cluster.  The figure shows a
$2\degr\times 2\degr$ region centered on Praesepe, which agrees well
with the cluster core radius of $\sim 1\degr$ \citep{Adams2002}. Also
indicated on the figure is the region of sky studied by
\citet{Claver2001}.  All of the cluster member WDs are located within
$\sim 35\arcmin$ of the cluster center, even though the region studied
in \citet{Claver2001}
extends $\gtrsim 50\arcmin$ out from the center.

\begin{figure*}
\plottwo{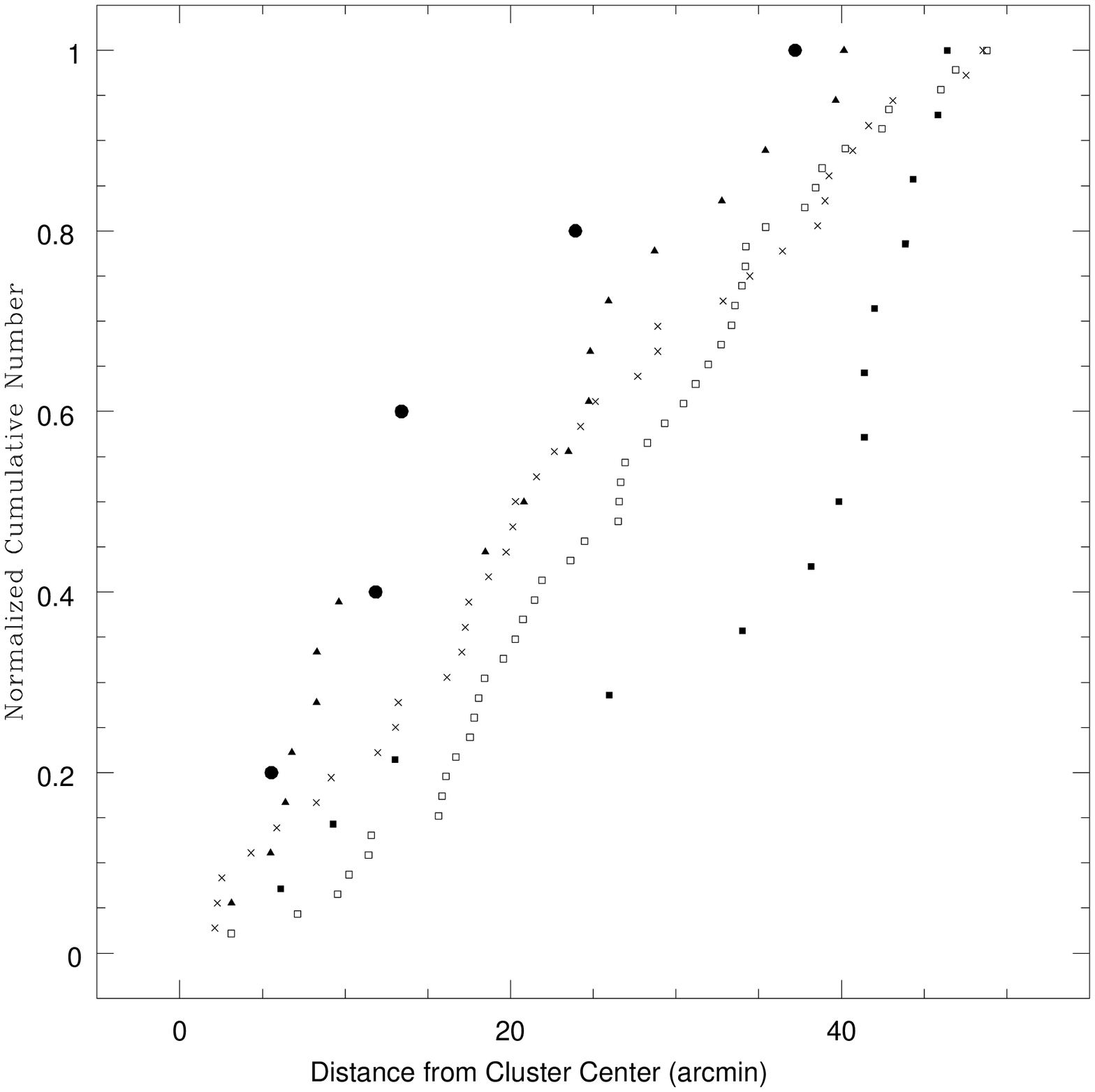}{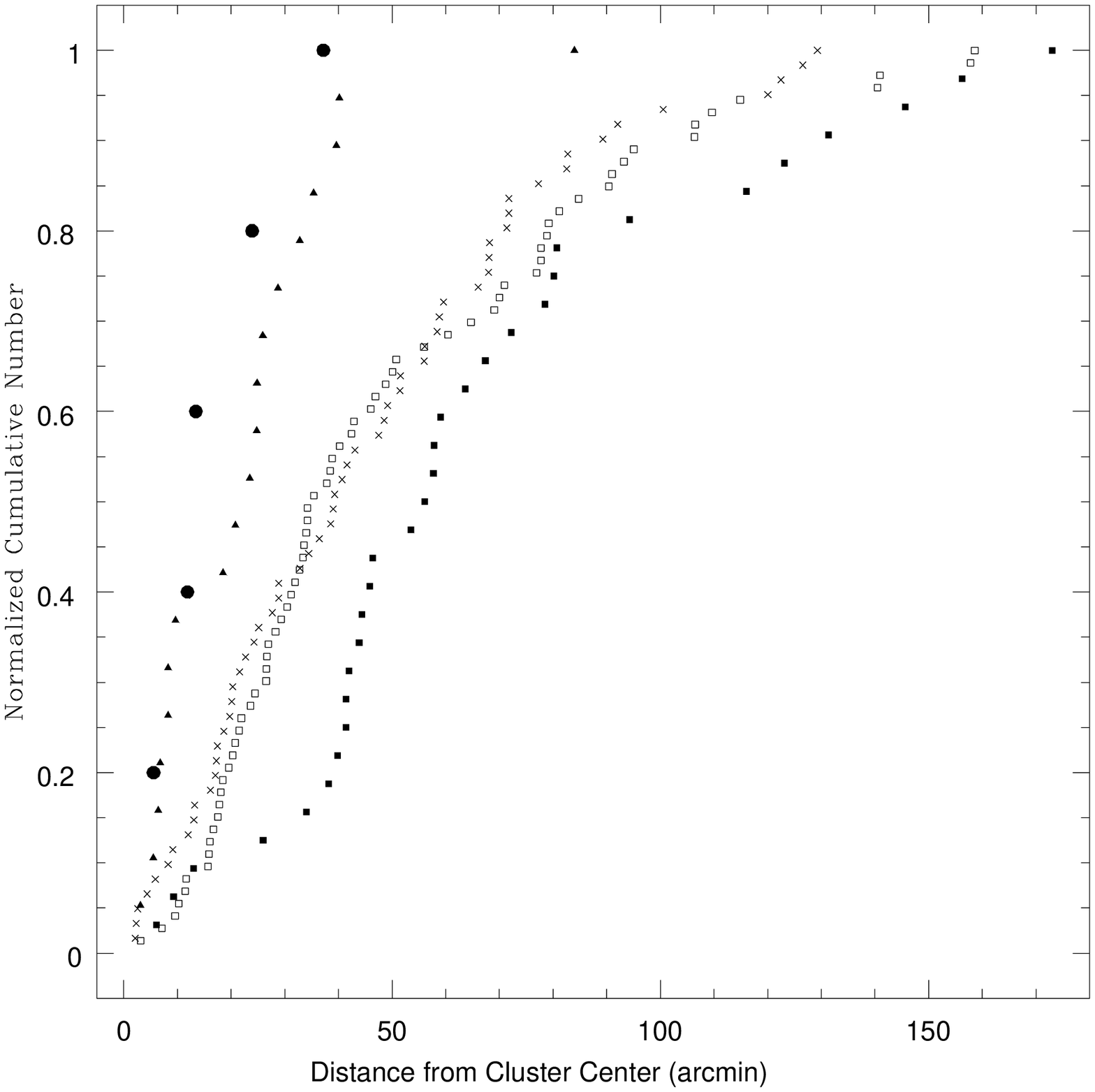}
\caption{Radial distribution of stars and WDs in Praesepe (a) interior to
  $50\arcmin$ and (b) over the entire cluster.
Large, filled circles show the cumulative radial profile of the five
known WDs. The remaining data are taken from \citet{Raboud1998b}, show
the profile of stars with $M > 2.5\msun$ (triangles), $2.5\msun \geq M
> 1.6\msun$ (crosses), $1.6\msun \geq M > 1\msun$ (open squares), $M
\leq 1\msun$ (filled squares).  The zero-age main-sequence masses of
Praesepe giants are $\approx 2.7\msun$. \label{praesepe_dist.fig}}
\end{figure*}

Figure~\ref{praesepe_dist.fig}a compares the cumulative
distribution of WDs as a function of radius from the cluster center to
the distribution of cluster stars interior to $50\arcmin$ 
from data by \citet{Raboud1998b}.  From the figure it appears that the WD
distribution follows that of the massive cluster stars, exactly as
predicted for a WD population with no net velocity kick.  A K-S test
comparing the WD radial distribution to that of the stars finds that
the WD distribution is consistent with that of the stars with masses
$\gtrsim 1.6\msun$ and weakly inconsistent with the distribution of
lower mass stars. 

To test further the significance of the WD radial distribution, two 
Monte Carlo simulation were performed.  In the first, 
five stars were drawn from the central $50\arcmin$ radial distribution
of the $\leq 1.2\msun$ stars in \citet{Raboud1998b}. Out of 10,000
realizations, 13\% had all five selected stars interior to the
$35\arcmin$ extent of the known WD population. The second Monte Carlo
simulation drew stars from the radial distribution of $M\leq 1.2\msun$
stars from the entire \citet{Raboud1998b} sample until five stars
interior to $35\arcmin$ were found. Of these, only 15\% of the
realizations were found to have no additional stars in the annulus
between $35\arcmin$ and $50\arcmin$. 

These simulations hint that the WD population is more
consistent with the giants and subgiants of Praesepe than with the
lower mass, as-yet unevolved cluster members, as predicted by the
N-body calculations that do not include a velocity kick.
Figure~\ref{praesepe_dist.fig}b shows the resulting radial
distribution of WDs compared with those for other Praesepe stars if
the five known WDs represent the entire WD population of Praesepe.
In this case, the similarity of the WDs with subgiant and giant
Praesepe members is even more striking.

The concentration of WDs toward the center of the cluster may not be
inconsistent with the concept of velocity kicks if all the WDs with
non-zero kicks have already been lost from the cluster (or are at
larger radii).  It would be useful for N-body simulations of 
open clusters to explore how the apparent radial distribution of WDs
varies with the strength of the applied kick.

We also note that the Pleiad WD (\objectname{LB 1497}) is located $\sim
1\fdg 4$ away from the tightly-concentrated core of the cluster, well
outside the half-mass radius ($\approx 53\arcmin$) but within the
Pleiades tidal radius ($\approx 7\,\fdg 4$) \citep{Raboud1998}.  This
contrasts with the central concentration of WDs observed in Praesepe
and further serves to muddy the waters.

\subsubsection{Shape of the Initial-Mass Function}
It is also possible that the dearth of WDs in Praesepe could be due to
a steeper IMF than normally considered.  To explore this possibility,
we use the code from \citet{Williams2004} to calculate the WD
population of Praesepe for various IMF slopes.  
The parameters used to normalize the IMF and
the WD detection criteria are the same as those for Praesepe presented
in \citet{Williams2004}; $\mcrit$ is taken to be $8\msun$, and the
binary fraction is taken to be 0.4.  Two different IMFs were explored
-- a power-law IMF of the form $\xi(M) \propto M^{-(1+\Gamma)}$ and
the broken power-law IMF from \citet{Naylor2002}, with a steep
high-mass power law IMF ($\Gamma=\Gamma_1$ for $M> 1\msun$) and a flat
low-mass slope ($\Gamma=\Gamma_2=0.2$ for $M\leq 1\msun$).

The Praesepe member \objectname{WD 0837+199} has a progenitor mass
$\leq 4.17\msun$ \citep{Claver2001}. We therefore calculate the
likelihood that, for a given IMF slope $\Gamma$, no WDs with
progenitor masses $\geq 4.17\msun$ are observed.  The value of
$\Gamma$ was then varied to find the slopes corresponding to
confidence levels of 60\%, 90\%, and 95\%.  Results are given in Table
\ref{calculations.tab}.  The calculations show that the lack of
high-progenitor-mass WDs, the lack of WDs observed in binary systems,
and the overall deficit of WDs in Praesepe can be explained by a
steep high-mass  IMF. Assuming Poisson statistics, both the
$\Gamma=2.55$ power-law IMF and the $\Gamma_1=2.75$ broken power-law
IMF are remarkably consistent with the Praesepe WD population, though
such steep IMFs are contrary to what is observed in younger open
clusters \citep{Kroupa2002,Kroupa2003}.

\begin{deluxetable*}{lcccccccc}
\tablecolumns{8}
\tablecaption{Calculation of the Praesepe WD population
  \label{calculations.tab}} 
\tablewidth{0pt}
\tablehead{& & & \multicolumn{5}{c}{$N_{WD}$} \\ \cline{4-8}
    & &            &       &          & Observed & Total &
  Observed \\
IMF\tablenotemark{a} & Confidence & $\Gamma$ & Total & Observed & in Binaries &
($M_i>4.17\msun$) & ($M_i > 4.17\msun$)}
\startdata
P & 60\% & 2.55 &  5.7 & 4.1 & 1.6 & 1.9 & 0.95 \\ 
P & 90\% & 2.02 & 12.1 & 8.0 & 2.8 & 4.8 & 2.3 \\
P & 95\% & 1.87 & 15.0 & 9.7 & 3.3 & 6.2 & 3.0 \\
N & 60\% & 2.75 &  6.4 & 3.4 & 0.7 & 2.1 & 0.9 \\
N & 90\% & 2.16 & 13.6 & 7.2 & 1.3 & 5.1 & 2.3 \\
N & 95\% & 2.75 & 16.7 & 8.8 & 1.6 & 6.6 & 3.0 \\
\enddata
\tablenotetext{a}{P= Power-Law IMF, N = \citet{Naylor2002} IMF}
\end{deluxetable*}

The present-day mass function (PDMF) of Praesepe has been discussed by
several authors.   \citet{Adams2002} find a PDMF slope of $\Gamma=0.6$
for $1\msun\geq M \geq 0.4\msun$, somewhat shallower than the Salpeter
slope of $\Gamma=1.35$ and far shallower than the IMFs needed to
explain the WD deficit. \citet{Williams1995} find a steeper IMF for
stars more massive than $0.6\msun$, with $\Gamma=1.7$, though this is
still shallower than the IMF required by our calculations. 

\subsection{Future work}

In order to determine the significance of the observed WD deficit in
Praesepe, and therefore determine the validity of the above
discussion, it is necessary to conduct a search for Praesepe WDs out
to larger radii.  Including proper motions in such a study would be
very helpful in reducing the large number of background and
extragalactic objects. Unfortunately, existing proper motion surveys
of Praesepe do not extend quite faint enough \citep[e.g.][]{Jones91}.

Examination of the complete WD content of additional clusters should
be able to constrain these hypotheses even further.  Comparison of the
WD populations of similar-age clusters of varying total masses can
determine the degree of any dynamical  contributions to the WD
deficit, as more massive clusters should retain a larger fraction of
WDs if dynamical effects dominate the WD deficit.   Examination of
young stellar clusters with masses similar to Praesepe and the Hyades
may help to constrain the high-mass IMFs, especially if a steep IMF is
observed in these clusters.

\acknowledgements This research is funded by NSF grants AST-0307492
and AST-0307321. The  authors wish to thank T.~von Hippel for helpful
comments on this manuscript and the anonymous referee for helpful
comments improving this paper.  The Digitized Sky Surveys were produced
at the Space Telescope Science Institute under U.S. Government grant
NAG W-2166. The images of these surveys are based on photographic data
obtained using the Oschin Schmidt Telescope on Palomar Mountain and
the UK Schmidt Telescope. The plates were processed into the present
compressed digital form with the permission of these institutions.
This research has made use of the SIMBAD database, operated at CDS,
Strasbourg, France.  Some of the data presented herein were  obtained
at the  W.M. Keck Observatory, which is operated as a scientific
partnership among the California Institute of Technology, the
University of California and the National Aeronautics and Space
Administration. The Observatory was made possible by the generous
financial support of the W.M. Keck Foundation. The authors wish to
recognize and acknowledge the very significant cultural role and
reverence that the summit of Mauna Kea has always had within the
indigenous Hawaiian community.  We are most fortunate to have the
opportunity to conduct observations from this mountain.


\begin{thebibliography}{}

\bibitem[Adams et al.(2002)]{Adams2002} Adams, J.~D., Stauffer, J.~R.,
Skrutskie, M.~F., Monet, D.~G., Portegies Zwart, S.~F., Janes, K.~A.,
\&  Beichman, C.~A.\ 2002, \aj, 124, 1570
\bibitem[Baumgardt \& Makino(2003)]{Baumgardt2003} Baumgardt, H.~\&
Makino,  J.\ 2003, \mnras, 340, 227
\bibitem[CLBK01()]{Claver2001} Claver, C.~F., Liebert, J.,  Bergeron,
P., \& Koester, D.\ 2001, \apj, 563, 987 (CLBK01)
\bibitem[Fellhauer et al.(2003)]{Fellhauer2003} Fellhauer, M., Lin,
D.~N.~C., Bolte, M., Aarseth, S.~J., \& Williams, K.~A.\ 2003, \apjl,
595,  L53
\bibitem[Gould \& Kollmeier(2004)]{Gould2004} Gould, A.~\& Kollmeier,
  J.A.\ 2004, \apj, in press
\bibitem[Jones \& Stauffer(1991)]{Jones91} Jones, B.~F.~\& Stauffer, 
J.~R.\ 1991, \aj, 102, 1080 
\bibitem[Koester \& Reimers(1996)]{Koester1996} Koester, D.~\&
Reimers, D.\  1996, \aap, 313, 810
\bibitem[Kroupa(2001)]{Kroupa2001} Kroupa, P.\ 2001, \mnras, 322, 231
\bibitem[Kroupa(2002)]{Kroupa2002} Kroupa, P.\ 2002, Science, 295, 82
\bibitem[Kroupa \& Weidner(2003)]{Kroupa2003} Kroupa, P.~\& Weidner,
C.\  2003, \apj, 598, 1076
\bibitem[Naylor et al.(2002)]{Naylor2002} Naylor, T., Totten,  E.~J.,
Jeffries, R.~D., Pozzo, M., Devey, C.~R., \& Thompson, S.~A.\ 2002,
\mnras, 335, 291
\bibitem[Oke et al.(1995)]{Oke1995} Oke, J.~B., et al.\ 1995, \pasp,
107,  375
\bibitem[Perryman et al.(1998)]{Perryman1998} Perryman, M.~A.~C., et
al.\  1998, \aap, 331, 81
\bibitem[Portegies Zwart et al.(2001)]{Portegies Zwart2001} Portegies
Zwart, S.~F., McMillan, S.~L.~W., Hut, P., \& Makino, J.\ 2001,
\mnras,  321, 199
\bibitem[Raboud \& Mermilliod(1998a)]{Raboud1998} Raboud, D.~\&
Mermilliod,  J.-C.\ 1998, \aap, 329, 101
\bibitem[Raboud \& Mermilliod(1998b)]{Raboud1998b} Raboud, D.~\&
Mermilliod,  J.-C.\ 1998, \aap, 333, 897
\bibitem[Richer et al.(1998)]{Richer98} Richer, H.~B., Fahlman, G.~G., 
Rosvick, J., \& Ibata, R.\ 1998, \apjl, 504, L91 
\bibitem[Salpeter(1955)]{Salpeter1955} Salpeter, E.~E.\ 1955, \apj,
121, 161
\bibitem[Vanden Berk et al.(2001)]{Vanden Berk2001} Vanden Berk,
D.~E., et  al.\ 2001, \aj, 122, 549
\bibitem[Wagner et al.(1986)]{Wagner1986} Wagner, R.~M., Starrfield,
S.~G.,  Sion, E.~M., Liebert, J., \& Nataliazotov 1986, \pasp, 98, 552
\bibitem[Weidemann et al.(1992)]{Weidemann1992} Weidemann, V., Jordan,
S.,  Iben, I.~J., \& Casertano, S.\ 1992, \aj, 104, 1876
\bibitem[Williams, Rieke, \& Stauffer(1995)]{Williams1995} Williams,
D.~M.,  Rieke, G.~H., \& Stauffer, J.~R.\ 1995, \apj, 445, 359
\bibitem[W04()]{Williams2004} Williams, K.~A.\ 2004, \apj, 601,  1067
(W04)
\end{thebibliography}
\end{document}